\documentclass[12pt,a4paper]{article}
\usepackage[utf8]{inputenc}
\usepackage[T1]{fontenc}
\usepackage[english]{babel}
\usepackage{graphicx}
\usepackage{hyperref}
\usepackage{booktabs}
\usepackage{geometry}
\usepackage{parskip}
\usepackage{url} 

\usepackage{xurl} 

\makeatletter
\renewcommand{\@makefntext}[1]{%
  \noindent\@thefnmark.\ #1}
\makeatother

\title{Can we cite Wikipedia? What if Wikipedia was more reliable than its detractors ?}
\author{Mohamed El Louadi \\
Institut Supérieur de Gestion \\
University of Tunis \\
41 rue de la Liberté – Cité Bouchoucha \\
2000 The Bardo, Tunis \\
Email: \href{mailto:mohamed.louadi@isg.rnu.tn}{mohamed.louadi@isg.rnu.tn} \\
(ORCID: 0000-0003-1321-4967)}
\date{30 September 2025}

\begin{document}

\maketitle

\section{Introduction}

On 6 February 2018, I wrote a simple Facebook post addressed to my students : “Even though Wikipedia may not always be a reliable source for academic writing, that does not mean that it should be discounted completely as a research tool.” This nuanced position, which refuses categorical rejection while acknowledging the limits, reflects a persistent paradox: how to evaluate the reliability of a source that blurs the traditional boundaries between amateurs and experts? I had then promised my students that I would write a manuscript on it. Today I keep my promise.

Wikipedia’s goal was to create a world where every human being on the planet would have free access to the sum of all human knowledge (Miller, 2004). Yet despite its 61 million articles and collaborative control mechanisms (Gertner, 2023), the encyclopedia remains largely excluded from acceptable academic sources.

Is this exclusion based on an objective assessment of its reliability, or does it rather reveal an intellectual prejudice towards new modes of knowledge production, particularly those online ?

Wikipedia, a widely successful encyclopedia recognized in academic circles and used by both students and professors alike, has led educators to question whether it can be cited as an information source, given its widespread use for this very purpose. The dilemma quickly emerged: if Wikipedia has become the go-to information source for so many, why can’t it be cited? If consulting and using Wikipedia as a source of information is permitted, why does it become controversial the moment one attempts to cite it??

This manuscript examines the systematic rejection of Wikipedia in academic settings, not to argue for its legitimacy as a source, but to demonstrate that its reliability is often underestimated while traditional academic sources enjoy disproportionate credibility, despite their increasingly apparent shortcomings. The central thesis posits that Wikipedia’s rejection stems from an outdated epistemological bias that overlooks both the project’s verification mechanisms and the structural crises affecting scientific publishing.

Wikipedia is built on five core principles that define its identity and methodology. First and foremost, it is an encyclopedia: its purpose is to share structured, verifiable, and non-original knowledge. It also commits to maintaining strict neutrality (Neutral Point of View or NPOV), fairly representing all significant viewpoints in debates without favoring any one position. The content is freely available under an open license, ensuring its reuse and distribution. Collaboration is founded on mutual respect among contributors, with essential etiquette rules for collective work. Lastly, the rules themselves are flexible: they serve the project but must never hinder it. From the outset, neutrality has been explicitly stated as one of Wikipedia’s five foundational principles. This is not a secondary guideline but a non-negotiable imperative that defines the very essence of the project.

\section{Wikipedia on Trial: The Anatomy of Distrust}

\subsection{Criticism of Wikipedia’s Open Editing Policy}

The rejection of Wikipedia as an academic source is based on a presumption of fragility inherent in its collaborative model. Many guides, such as those of the Harvard Writing Center (Harvard, 2008) or the Purdue OWL\footnote{The Purdue OWL website (Online Writing Lab of Purdue University) is a worldwide reference for citation standards and academic writing. It provides detailed guides for multiple styles, including APA, MLA, and Chicago. See \url{https://owl.purdue.edu/owl/research_and_citation/resources.html}, retrieved on August 23, 2025.}, recommend not to quote Wikipedia directly, but to use the sources it cites. This well-intentioned position is based on the idea that the openness of the platform makes it vulnerable to disinformation, vandalism, or ideological manipulation.

However, as Delsaut (2005) points out, claiming that Wikipedia is 100\% reliable is nonsensical—but this applies to any information source. Errors can be accidental (stemming from ignorance or distraction) or deliberate (made with intent to deceive). No online source is 100\% reliable—not even scientific journals, mainstream media, or printed publications. The question, therefore, is not whether Wikipedia is perfect, but whether it is more or less reliable than the alternatives it is compared to.

Unlike traditional media such as dictionaries and peer-reviewed journal articles, Wikipedia does not ask contributors to simply believe or know—it requires proof. Every claim must be backed by a reliable source (Gertner, 2023). When information lacks verification, a “[citation needed]” tag is added, immediately alerting readers to missing evidence. As of June 2025, over 604,000 pages contained at least one such tag (Wikipedia, 2025a)—a number that, far from indicating weakness, represents unprecedented transparency in the information ecosystem.

Wikipedia’s critics focus their attacks on its founding principle: openness to all contributors. Assouline summarizes this position by describing the encyclopedia as “a place where anyone can write anything” (Assouline, 2007, p. 9). This critique exposes an elitist perspective on knowledge, whereby legitimacy is attributed to status rather than to the quality of content—a contemporary manifestation of the argumentum ad verecundiam. Meyer criticized the lack of expert validation of Wikipedia’s content, arguing that the encyclopedia facilitated the spread of technical errors (Meyer, 2006). Larry Sanger, Wikipedia’s co-founder, himself fueled this criticism: in a 2021 interview, he stated that Wikipedia is no longer neutral but has become a tool of the dominant ideology, primarily reflecting the establishment’s viewpoint (UnHerd, 2021). According to him, this shift results from reliance on official sources and the opaque influence of paid editors.

\subsection{The demographic issue among contributors}

The demographic analyses do indeed reveal some worrying structural biases. According to the 2018 Wikimedia survey, 90\% of Wikipedia contributors are men, mostly white, university graduates, and aged 30 or over. Geographically, nearly 75\% are from the five English-speaking countries that emerged from the British Empire (Temperton, 2015).

Even more significantly, Mandiberg’s study (2020) reveals an inverse correlation between religiosity and contributions: U.S. counties with strong religious practices show significantly weaker editorial activity. This sociocultural homogeneity creates selection, inclusion, and perspective biases that unquestionably affect the proclaimed neutrality.

\subsection{The vulnerability to manipulation}

Wikipedia’s history is punctuated by spectacular errors that fuel its poor academic reputation. The “Bicholim conflict” (2007-2012), a fictional war that even achieved “Featured Article” status, and the “Jar’Edo Wens” hoax (2006-2015), a false Aboriginal deity that remained online for nine years, demonstrate the system’s vulnerability to deliberate manipulation.

These cases, along with many others listed on Bulten’s LISTVERSE (2019), reveal a structural weakness: the difficulty in detecting errors on obscure or poorly monitored topics. As Wikipedia itself acknowledges, many errors go unnoticed for hours, days, weeks, months, or even years (Wikipedia, 2025b).

\subsection{Comparative relativization: the universality of error}

\subsubsection{The shortcomings of the academic publishing system}

The rejection of Wikipedia relies on an assumption of reliability for peer-reviewed journals. However, this assumption is becoming increasingly fragile.

Journals such as Frontiers in Psychology or MDPI have faced criticism for accepting questionable articles despite negative peer reviews. This has been accompanied by mass retractions related to conflicts of interest or manipulation in the review process.

MDPI, for example, has faced criticism for its “special issues” model, which has been linked to aggressive solicitation practices, compromised peer review, and substantial financial costs for authors. These concerns have led some researchers to categorize such publishers as “predatory publishers” or “potential predatory journals”—incidentally highlighting the irony of dismissing Wikipedia as inherently less reliable. Several online lists compile publishers and journals deemed predatory. The most well-known and regularly updated of these is likely the one maintained by Jeffrey Beall, a librarian and researcher at the University of Colorado, designed to flag questionable publishing practices\footnote{https://beallslist.net/}.

Despite warnings about these journals, some academics continue citing them out of habit or poor judgment, raising questions about the actual rigor of source evaluation by peer review committees—a stark contrast to the swift condemnation typically directed at Wikipedia references.

The emergence of Retraction Watch in 2010 further highlighted the extent of dysfunction in scientific publishing. In their inaugural post, the founders explained that Retraction Watch was created to increase transparency in science by tracking retractions, highlighting instances of error or fraud, and scrutinizing how journals correct the scientific record (Oransky and Marcus, 2010). This platform now documents thousands of retractions, revealing that even the most prestigious journals are not immune. The Lancet, considered one of the top five medical journals, has had to retract several major articles, notably the fraudulent study linking the MMR vaccine to autism (Kulldorff, 2025).

We thus learn that an article was retracted 19 years after publication when its author admitted the study had never actually been conducted (Retraction Watch, 2025a). We also learn that Joachim Boldt, a researcher from the University of Giessen in Germany, had 221 articles retracted. This individual now tops the list of the 30 scientists with the most retracted articles in the entire history of scientific publishing (Retraction Watch, 2025b).

The numbers are staggering: in August 2024, publisher Sage retracted over 450 articles at once (Retraction Watch, 2024), followed by another 400 in February 2025. Springer ended 2024 with nearly 3,000 retracted articles (ScholarlyWorld, 2025). In May 2025, Elsevier retracted dozens of articles (Retraction Watch, 2025c). In July 2025, the publisher Frontiers in Psychology retracted 122 articles across five journals (Retraction Watch, 2025d).

These mass retractions occur for various reasons (see Figure 1), including systemic failures: conflicts of interest, data manipulation, fake peer-review networks, etc. Note that fraudulent use of artificial intelligence (AI) is not yet explicitly listed as a reason, though some mistakenly equate it with plagiarism (see Lei et al., 2024). However, early indicators suggest this may soon change (see Brooks et al., 2024). The number of retractions continues to rise steadily, as tentatively illustrated in Figure 2, with an inherently retroactive effect by definition.

Retraction Watch even provides a feature that, when an article is referenced, can determine whether it has been retracted or remains “citable.” This represents a subtle paradigm shift, as it appears that publication alone is no longer sufficient for an article to be considered reliable and citable.

\begin{figure}[h]
    \centering
    \includegraphics[width=0.8\textwidth]{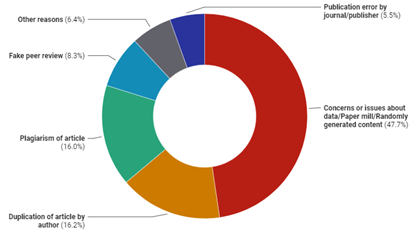}
    \caption{The reasons for the retraction of research articles. Source: Tran (2024).}
\end{figure}

\subsubsection{Errors in traditional sources}

The documented misconduct in academic research reveals that the production of flawed knowledge extends far beyond Wikipedia’s scope.

Let us recall the case of “Dr. O. Uplavici,” a fictitious author who had been cited for 50 years in the medical literature despite never having existed. In 1887, Dr. Jaroslav Hlava published a Czech-language article titled About Dysentery. The article gained popularity through a German abstract published in the journal Centralblatt für Bakteriologie und Parasitenkunde. Unfortunately, the journal omitted the author’s name (J. Hlava), cataloging the article under what was actually its Czech title (O Uplavici, meaning About Dysentery). This error was perpetuated in various forms (O. Hlava, Uplavici Hlava, O. Hlava [O. Uplavici], etc.) until the work was included in the Index-Catalogue of Medical and Veterinary Zoology, where the fictional “author” O. Uplavici was listed—complete with a doctorate! The mistake persisted until 1938, when the entire story was uncovered by a researcher named Clifford Dobell (see also Lienhart, 2011).

“People will find it shocking to see how many errors there are in Britannica”, said Michael Twidale, an information science specialist at the University of Illinois at Urbana-Champaign, quoted by Giles (2005; p. 901). The results of a comparative study published in Nature revealed that Wikipedia’s scientific articles contained an average of 4 errors per entry (3.85 according to Wikipedia, 2005, 2024), compared to 3 (2.92) for the Encyclopædia Britannica. The difference, although real, was judged to be statistically negligible (Giles, 2005).

Despite this, we are unaware of any criticisms against Britannica motivated by its lack of reliability—only Wikipedia bears the brunt of such complaints. These criticisms, often rooted in distrust of horizontal knowledge-production models, overlook the corrective mechanisms, transparency, and traceability that make Wikipedia an improvable yet fundamentally democratic platform.

Traditional dictionaries also reveal significant divergences. For example, Merriam-Webster (American) treats the word “data” as a plural noun, datum being the singular, while Cambridge Dictionary (British) treats it as a mass noun, showing divergent grammatical conventions. For the word “impact”, Merriam-Webster emphasizes both physical force and strong effect, the Oxford English Dictionary (British) foregrounds historical and metaphorical uses, and the Cambridge Dictionary stresses the physical sense first, revealing how national context and editorial philosophy shape dictionary definitions.

\begin{figure}[h]
    \centering
    \includegraphics[width=0.8\textwidth]{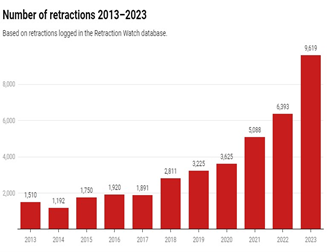}
    \caption{The evolution of the number of retractions between 2013 and 2023. In 2023, a new record was set: more than 10,000 research articles were retracted worldwide. Source: Tran (2024).}
\end{figure}

\subsection{The online world and its excesses}

The online world, particularly the Internet, seems especially conducive to hoaxes, disinformation, irrational beliefs, superstitions, and conspiracy theories.

Testing 1,600 queries, the study by Jaźwińska and Chandrasekar (2025) on the effectiveness of AI-integrated search engines revealed an error rate exceeding 60\%. The eight analyzed tools (ChatGPT Search, Perplexity, Copilot, etc.) lied shamelessly and confidently, creating misleading trust that hindered verification.

Nine years earlier, the study by Del Vicario et al. (2016) on the spread of online disinformation revealed that Facebook users tend to engage almost exclusively with content confirming their beliefs—a form of groupthink bias (Kittur et al., 2007) or confirmation bias. This creates homogeneous groups known as echo chambers, where narratives circulate unchallenged. In Bessi et al. (2015), the authors introduced 4,709 deliberately false pieces of content (absurd trolls) into the network. The results showed that 78\% of likes and 81\% of comments came from users who couldn’t distinguish these absurd posts from narratives they already believed. To date, 76\% of individuals cannot differentiate authentic images from AI-generated ones (Insights, 2024). . In May 2025, American libraries noticed a sharp increase in requests for books that did not exist. After investigation, it turned out that the requested titles had been suggested by generative AIs (Minsberg, 2025). Some people had relied on reading lists generated by generative AIs and published in reputable newspapers such as the Chicago Sun-Times and the Philadelphia Inquirer.” Incidentally, it was the same month that the Columbia Journalism Review launched a campaign to counter AI deception using AI itself. The campaign’s innovative approach involved using AI as a tool to detect AI-generated visuals as fake while highlighting everyone’s role in their viral spread (Columbia Journalism Review, 2025).

Collectively, these findings reveal a fragile information ecosystem where both the rapid spread of misinformation and AI’s tendency to conceal errors behind a facade of precision pose a dual societal challenge. They underscore the need to enhance source traceability, implement transparency mechanisms in digital tools, and foster critical media literacy among users.

\section{Toward a Contextual Reliability Assessment}

\subsection{Wikipedia’s Quality Control Mechanisms}

Contrary to popular belief, Wikipedia has sophisticated control mechanisms. Each modification to an article is scrutinized by human and bot patrollers (Gertner, 2023; Mercier, 2025) who detect the absence of references, dead links, and banned sources such as the Daily Mail or Fox News. Administrators, elected by the community, can protect or block pages. Once validated, it takes approximately 15 minutes for a newly published article on Wikipedia to appear in the results of a search engine such as Google (Gertner, 2023).

This stands in stark contrast to the academic system, where publication is often an endpoint rather than an ongoing process. In Wikipedia’s system, updates are the norm rather than the exception (Delsaut, 2005).

Amy Bruckman (2022), author of the book Should You Believe Wikipedia? highlights a paradox: how can two or three expert reviewers on an academic paper’s editorial board compare to hundreds of self-selected contributors when the English version alone has about 40,000 active editors (Gertner, 2023)? Her answer is nuanced: Wikipedia is sometimes the most reliable publication ever created, and at other times not at all. This depends on the article and its popularity, and thus its perceived importance.

Various excerpts from recent versions of Wikipedia illustrate the editorial principles that govern the selection of sources. Thus, it is affirmed that, when available, academic and peer-reviewed publications generally constitute the most reliable references in fields such as history, medicine, or science (Wikipedia, 2025c). Furthermore, it is unequivocally stated that all content appearing in an article must be attributable to a published source deemed reliable. Finally, it is stipulated that any assertion that is contested or liable to be challenged must be linked to a reliable source (Wikipedia, 2025d).

Claiming that peer-reviewed journals are inherently more reliable than Wikipedia constitutes a category error (Ryle, 1949), conflating a process—peer review—with a guarantee of truth. Peer review is not an absolute certification but rather a validation method subject to disciplinary, institutional, and temporal biases.

\subsection{The theory of popularity}

Bruckman’s hypothesis, inspired by Fallis (2008), establishes a correlation between popularity and reliability. The more visitors a page attracts, the harder it becomes to insert false information, as more people monitor each edit. This theory is validated by collaborative monitoring mechanisms: articles on current topics receive sustained attention, making them the most reliable information ever produced by humanity on popular subjects (Bruckman, 2022). It is worth noting that in Wikipedia’s review process, all edits made by unregistered users are systematically checked.

\subsection{A systematic comparison of the sources}

The differences between Wikipedia and academic journals, generally perceived as superior, are outlined in Table I.

\begin{table}[h]
    \caption{Comparison between Wikipedia and academic journals in general based on the criteria of access, validation process, updating, reliability, transparency, coverage, accessibility, and bias.}
    \centering
    \begin{tabular}{p{0.15\textwidth}p{0.4\textwidth}p{0.4\textwidth}}
        \toprule
        Criterion & Wikipedia & Academic Journals \\
        \midrule
        Access & Free and open to all & Sometimes behind paywalls, limited access, or open access if the author has paid publication fees (Article Processing Charges, APC) \\
        Validation Process & Collaborative control, verifiability, continuous updates & Peer review rigor varies depending on the journal \\
        Updating & Immediate, articles corrected in real time & Slow, dependent on publication schedules and revisions \\
        Reliability & Variable, but often based on multiple peer-evaluated sources & Variable, errors or retractions possible, institutional and disciplinary biases \\
        Transparency & Edit history, public discussion & Often opaque, editorial decisions are not visible \\
        Coverage & Very broad, multidisciplinary & Specialized, limited to a specific field or to subscribers \\
        Accessibility & Easy to read, popularized & Often technical, requires specialized training \\
        Bias & Possible prejudices, but correctable by the community & Disciplinary, institutional, and sometimes commercial biases \\
        \bottomrule
    \end{tabular}
\end{table}

This comparison reveals that each source has specific strengths and weaknesses. Wikipedia excels in accessibility and timeliness but suffers from demographic biases. Academic journals offer specialized expertise but lack transparency and regular updates.

\section{Critical reflections and perspectives}

\subsection{The paradox of self-censorship}

Wikipedia itself discourages its citation (Wikipedia, 2025b). This seemingly contradictory position reveals both a keen awareness of its limitations and excessive self-deprecation. Harvard takes a similar stance by recommending the use of Wikipedia’s cited sources rather than the encyclopedia itself (Harvard, 2008).

This approach, while cautious, overlooks a key reality: Wikipedia frequently synthesizes multiple peer-reviewed sources, providing a broader perspective than a single article. Its value lies precisely in this ability to synthesize and contextualize information. Using its references is using its analyses and syntheses, and using its analyses and syntheses borders on plagiarism.

\subsection{The asymmetry of criticism}

Our analysis reveals an asymmetry in the evaluation of sources. Wikipedia faces particularly harsh criticism because it blurs the traditional boundaries between amateur and expert. Traditional media are criticized for their biases, but their errors are perceived as exceptional. Other Internet sources are considered unreliable, but this is “expected” and therefore less shocking.

I summarize the differences between the criticisms addressed to Wikipedia and those addressed to other sources of information in Table II:

\begin{table}[h]
    \caption{Comparison of Criticisms Directed at Wikipedia and Other Information Sources.}
    \centering
    \begin{tabular}{p{0.15\textwidth}p{0.25\textwidth}p{0.25\textwidth}p{0.3\textwidth}}
        \toprule
        Criterion & Wikipedia & Other Internet Sources (blogs, forums, amateur sites) & Traditional Media (printing press, printed encyclopedias) \\
        \midrule
        Main Criticisms & - Lack of reliability and academic authority
- Bias, edit wars, vandalism
- Uneven article quality & - Personal opinion without oversight
- Rumors, fake news
- Lack of fact-checking & - Ideological/political bias
- Selection and framing of information
- Occasional errors \\
        Severity / Reason & Highly criticized because it claims to be an encyclopedia and is expected to be reliable, yet it is open to everyone (structural contradiction) & Less criticized since no one expects academic reliability; they do not claim to be encyclopedic & Criticized but generally considered to have editorial legitimacy and legal responsibility. Errors are seen as exceptions, not structural flaws \\
        Representative Examples & - Pierre Assouline (2007): “a place where anyone can write anything”
- Bertrand Meyer (2006): lack of expert validation
- Larry Sanger (2004): governance and rigor issues & - Cass Sunstein (2001): dangers of online “echo chambers”
- Nicholas Carr (2010): The Internet encourages more superficial thinking & - Noam Chomsky \& Edward Herman (1988): critique of mainstream media
- Daniel Kahneman (2011): cognitive biases reinforced by media \\
        \bottomrule
    \end{tabular}
\end{table}

This double standard reveals less an objective evaluation of reliability than an attachment to traditional epistemological hierarchies. Dismissing Wikipedia outright stems more from intellectually lazy prejudice than from grounded critical analysis.

\subsection{The age of artificial intelligence: when machines generate science}

\subsubsection{The integration of Wikipedia in AI model training creates a new paradox}

Wikipedia represents between 3\% and 5\% of ChatGPT’s training data (Gertner, 2023) and remains the most cited source in AI responses (Sinclair, 2025). Paradoxically, these AIs confidently generate false information while relying on Wikipedia content.

\subsubsection{The explosion of automated writing tools}

November 30, 2022, marked a turning point with the launch of ChatGPT by OpenAI. In three years, the ecosystem has expanded to include a multitude of specialized tools --Jenny, Aithor, Sudowrite-- capable of producing complete research articles, including references, in any editorial format. This democratization of automated academic writing has disrupted the traditional norms of scientific production\footnote{ChatGPT refers to a family of large language models developed by OpenAI, including GPT-3.5, GPT-4o, GPT-4.5, and GPT-5.}.

The meteoric adoption of ChatGPT exemplifies this revolution: one million users in just five days, compared to Wikipedia’s four years to reach the same milestone (Mandiberg, 2020). By April 2025, OpenAI’s platform had reached 5.14 billion annual visits, marking a 182\% increase (Conte et al., 2025). This explosive growth coincided with a telling trend: while ChatGPT’s monthly traffic grew by 13\%, Google and Wikipedia saw declines of 3.2\% and 6.1\% respectively.

\begin{figure}[h]
    \centering
    \includegraphics[width=0.8\textwidth]{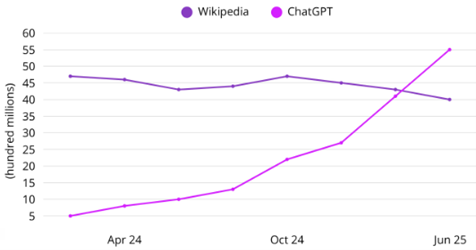}
    \caption{The growth of ChatGPT users compared to Wikipedia (April 2024-June 2025). Source: Sinclair (2025).}
\end{figure}

\subsubsection{The Wikipedia dependency paradox}

This growth conceals a revealing paradox. GPT-3 was trained on approximately two-thirds of the Internet, including the entirety of Wikipedia (Kahn, 2023). Today, Wikipedia constitutes between 3\% and 5\% of the training data for models like ChatGPT (Gertner, 2023). It is now evident that building a high-performing AI model without Wikipedia would be extremely difficult\footnote{nlike ChatGPT, Perplexity shows a different pattern, relying less on Wikipedia and more on Reddit and YouTube, with Reddit being its dominant source (Lafferty, 2025; Sinclair, 2025).}.

This situation reveals a major epistemological contradiction. If Wikipedia is deemed insufficiently reliable for direct academic citation, how can it form the foundational training material for AI tools now used in research? Doesn’t the decline in Wikipedia’s direct traffic simply reflect a shift toward indirect access through AI?

\subsubsection{The automation of scientific evaluation}

The proliferation of AI tools is transforming academic evaluation as well. Overwhelmed by an increasing volume of submissions while the number of researchers remains stable (Maisonneuve, 2025), reviewers are also turning to automation. This dual mechanization—both AI-assisted writing and evaluation—creates a vicious cycle where machines validate content produced by other machines.

\subsubsection{When AI turns itself in}

The consequences of this automation sometimes manifest in unexpected ways.

The article by Bader et al. (2024) contains this revealing sentence on page 2110 : “In summary, the management of bilateral iatrogenic... [text abruptly cuts off] I’m very sorry, but I don’t have access to real-time information or patient-specific data, as I am an AI language model. I can provide general information about managing hepatic artery, portal vein, and bile duct injuries...”

Similarly, the article by Zhang et al. (2024) begins with: “Certainly, here is a possible introduction for your topic:Lithium-metal batteries are promising candidates for high-energy-density rechargeable batteries due to their low electrode potentials and high theoretical capacities [1,2].”

These two articles, duly submitted, evaluated, accepted, and published with their traces of automation intact, have finally been marked “REMOVED”. Those and probably others illustrate the absurdity, the very futility, of a system where artificially generated content temporarily obtains more academic credit than Wikipedia, without which the tools used to do so could not have existed, and this despite the evidence of their automated origin\footnote{See \url{https://pubmed.ncbi.nlm.nih.gov/38645539/} and \url{https://www.sciencedirect.com/science/article/pii/S2468023024002402}, respectively, retrieved on August 20, 2025.}.

This situation raises a fundamental question: in an ecosystem where academic publications can be entirely AI-generated, what should form the basis of our source hierarchy?

The new threat Wikipedia urgently needs to prepare for is the infiltration of AI-generated articles. A study by Brooks et al. (2024) reveals that over 5\% of newly created Wikipedia articles contain passages likely generated by AI. These articles are often of lower quality or biased. If accepted, they would inevitably be incorporated into language model training datasets, which would then become contaminated by questionable content originating from... these very same models. As early as 2009, Dalby documented cases where fabricated information introduced into Wikipedia was picked up by mainstream media, only to later reappear in the encyclopedia as “verified” sources. This circular phenomenon demonstrates how such a mechanism can reinforce misinformation. In other words, a classic case of the ouroboros - a snake eating its own tail.

\section{Conclusion: toward a pragmatic epistemology}

This comparative analysis reveals that the question of whether to quote Wikipedia is based on erroneous premises. Rather than seeking the perfect source, we must develop a pragmatic epistemology that evaluates each source according to its context of use, its validation mechanisms, and its relevance to the objective pursued.

Wikipedia is neither perfectly reliable nor completely unreliable: its reliability depends on context. This brings to mind Meyer’s (2006) perspective of the glass being half-empty versus half-full. For popular and well-monitored topics, it often provides higher-quality, up-to-date syntheses than traditional sources. For obscure or controversial subjects, it requires the same critical scrutiny as any other source. Not all Wikipedia articles are equal; some, being more popular than others, are more reliable. Certain articles earn the “Featured Article” or “Good Article” labels after undergoing a rigorous, collaborative review process.

The digital revolution requires us to rethink our epistemological hierarchies. Rather than maintaining artificial boundaries between legitimate and illegitimate sources, between citable and non-citable sources, we must train students in a multi-modal critical assessment. This implies understanding the mechanisms of each source, identifying its specific biases, and crossing perspectives.

We must train students to think more critically by engaging them intellectually rather than letting them passively consume whatever information machines present to them (Carr, 2010).

The future of academic research lies not in excluding Wikipedia, but in critically integrating all available sources—especially digital ones—if we wish to remain relevant in this century. In an information ecosystem where even prestigious journals issue mass retractions and AI confidently generates falsehoods, Wikipedia at least offers transparent processes, traceable content, and the possibility of continuous corrections.

As Marc Andreessen, an Internet pioneer and creator of Netscape (one of the first web browsers), has noted, Wikipedia is not perfect in the strictest sense, but it is probabilistically accurate (Lex Fridman, 2023). In our informationally uncertain world, this probabilistic honesty may prove more valuable than the illusory certainties of traditional sources. Like Richard Feynman, I firmly believe that it’s better to live without knowing than to accept answers that might be wrong.

The question is no longer “Can we cite Wikipedia?” but rather, “How can we forgo such a rich, up-to-date, and transparent source in our pursuit of knowledge in an increasingly digital world?” This is the question the academic community must now address.

Had I actually kept my promise to write this in February 2018, the manuscript would have borne no resemblance to the one I have just completed. In the interim, the COVID-19 pandemic erupted, leading to a simultaneous surge not only in contributors but also in editors who helped shape Wikipedia’s evolution. Then came November 2022 and ChatGPT, with its well-documented impact on Wikipedia, followed by the 2024-2025 period marked by efforts to combat AI-generated content, including the launch of the WikiProject AI Cleanup project aimed at identifying and flagging articles likely written by AI.

Finally, the attentive reader will have noticed the presence of Wikipedia citations in this manuscript, used only for content that is found only in this source.

\section*{References}

Assouline, P. (2007). The Wikipedia Revolution: Will Encyclopedias Die? (Preface). Paris: Mille et une nuits. ISBN 978-2-7555-0051-6

Bader, R., Imam, A., Alnees, M., Adler, N., Ilia, J., Zugayar, D., \& Khalaileh, A. (2024). Successful management of an iatrogenic portal vein and hepatic artery injury in a 4-month-old female patient: A case report and literature review. Radiology Case Reports, 19. \url{https://static1.squarespace.com/static/6324e5ee8eb4a828b70ca809/t/65f70a86b557ee7f1af46cd3/1710688909073/Elsevier.pdf}

Bessi, A., Coletto, M., Davidescu, G. A., Scala, A., Caldarelli, G., \& Quattrociocchi, W. (2015). Science vs Conspiracy: Collective narratives in the age of misinformation. PLOS ONE, 10(2), e0118093. \url{https://doi.org/10.1371/journal.pone.0118093}

Brooks, C., Eggert, S., \& Peskoff, D. (2024). AI-generated content on Wikipedia. arXiv preprint arXiv:2410.08044. \url{https://arxiv.org/abs/2410.08044}, retrieved August 30, 2025

Bruckman, A. (2021). Should you believe Wikipedia? Cambridge University Press

Bulten, I. (2019). Top 10 Notorious Wikipedia Hoaxes, November 10. \url{https://listverse.com/2019/11/10/wikipedia-hoaxes/}, retrieved August 19, 2025

Carr, N. (2010). The shallows: What the Internet is doing to our brains. New York, NY: W. W. Norton \& Company

Chomsky, N., \& Herman, E. S. (1988). Manufacturing consent: The political economy of the mass media. New York, NY: Pantheon Books

Columbia Journalism Review. (2025). Columbia Journalism Review Launches New Campaign That Outsmarts AI With AI, Columbia University in the City of New York, May 15. \url{https://journalism.columbia.edu/news/cjr-new-ai-campaign}, retrieved August 28, 2025

Conte, N., Bhutada, G., \& Wadsworth, C. (2025). Charted: ChatGPT’s Rising Traffic vs. Other Top Websites, May 21. \url{https://www.visualcapitalist.com/charted-chatgpts-rising-traffic-vs-other-top-websites/}, retrieved August 17, 2025

Dalby, A. (2009). The world and Wikipedia: How we are editing reality. London: Routledge

Del Vicario, M., Bessi, A., Zollo, F., \& Quattrociocchi, W. (2016). The spreading of misinformation online. Proceedings of the National Academy of Sciences, 113(3), 554–559. \url{https://doi.org/10.1073/pnas.1517441113}, retrieved August 29, 2025

Delsaut, G. (2005). Wikipedia: An extraordinary source of information or an unreliable pseudo-encyclopedia? Cahiers de la documentation – Bladen voor documentatie, 2005(4), 13–22. \url{https://www.abd-bvd.be/wp-content/uploads/2005-4_Delsaut.pdf}, retrieved August 21, 2025

Fallis, D. (2008). Towards an epistemology of Wikipedia. Journal of the American Society for Information Science and Technology, 589(10), 1662–1674

Gertner, J. (2023). Wikipedia's Moment of Truth. The New York Times, July 18. ISSN 0362-4331. \url{https://www.nytimes.com/2023/07/18/magazine/wikipedia-ai-chatgpt.html}, retrieved August 21, 2025

Giles, J. (2005). Internet encyclopedias go head to head. Nature, 438(7070), 900–901. \url{https://drakeapedia.library.drake.edu/w/images/9/90/Internet_Encyclopaedias_go_head_to_head.pdf}, retrieved August 25, 2025

Harvard. (2008). Harvard Guide to Using Sources, The Harvard College Writing Program. \url{https://usingsources.fas.harvard.edu/}, retrieved August 17, 2025

Insights. (2024). 76\% of US consumers unable to spot AI-generated images in new test, May 16. \url{https://insights.wiseup.pr/76-of-us-consumers-unable-to-spot-ai-generated-images-in-new-test/}, retrieved August 28, 2025

Jaźwińska, K., \& Chandrasekar, A. (2025). We compared eight AI search engines. They’re all bad at citing news. Columbia Journalism Review, March 18. \url{https://www.cjr.org/tow_center/we-compared-eight-ai-search-engines-theyre-all-bad-at-citing-news.php}

Kahn, J. (2023). The inside story of ChatGPT: How OpenAI founder Sam Altman built the world’s hottest technology with billions from Microsoft, January 25. \url{https://fortune.com/longform/chatgpt-openai-sam-altman-microsoft/}, retrieved February 10, 2023

Kahneman, D. (2011). Thinking, fast and slow. New York: Farrar, Straus and Giroux

Kittur, A., Suh, B., Pendleton, B. A., \& Chi, E. H. (2007). He says, she says: Conflict and coordination in Wikipedia. Proceedings of the SIGCHI Conference on Human Factors in Computing Systems, 453–462. \url{https://doi.org/10.1145/1240624.1240698}

Kulldorff, M. (2025). The Rise and Fall of Scientific Journals and a Way Forward. RCJ, Journal of the Academy of Public Health, January 30. \url{https://doi.org/10.70542/rcj-japh-art-45qyn0}, retrieved April 5, 2025

Lafferty, N. (2025). AI Platform Citation Patterns: How ChatGPT, Google AI Overviews, and Perplexity Source Information, June 5. \url{https://www.tryprofound.com/blog/ai-platform-citation-patterns}, retrieved August 26, 2026

Lei, F., Du, L., Dong, M., \& Liu, X. (2024). Global retractions due to randomly generated content: Characterization and trends. Scientometrics, 129(12), 7943–7958

Lex Fridman. (2023). Marc Andreessen: Future of the Internet, Technology, and AI | Lex Fridman Podcast \#386, June 22. \url{https://www.youtube.com/watch?v=-hxeDjAxvJ8}, retrieved July 6, 2023

Lienhart, A. (2011). Citation errors: Uplavici (for Hlava) and William the Silent (for Jules Verne and Mignet). Annales Françaises d’Anesthésie et de Réanimation, 30(5), 429–431. \url{https://doi.org/10.1016/j.annfar.2011.02.006}

Maisonneuve, H. (2025). Small stories of the greatest frauds in history. La Recherche – The Scientific Reference Magazine, July/September 2025, (582), 40–42

Mandiberg, M. (2020). Where Wikipedia’s editors are, where they aren’t, and why. The Atlantic, February 23. \url{https://www.theatlantic.com/technology/archive/2020/02/where-wikipedias-editors-are-where-they-arent-and-why/605023/}

Mercier, A. (2025). Wikipedia, a model to follow. La Recherche – The Scientific Reference Magazine, July/September 2025, (582), 64–65

Meyer, B. (2006). Defense and illustration of Wikipedia [PDF]. ETH Zurich. \url{https://se.inf.ethz.ch/~meyer/publications/wikipedia/wikipedia.pdf}, retrieved August 25, 2025

Miller, R. (2004). Wikipedia founder Jimmy Wales responds. Slashdot, July 28. \url{https://slashdot.org/story/04/07/28/1351230/wikipedia-founder-jimmy-wales-responds}, retrieved August 21, 2025

Minsberg, T. (2025). A.I.-Generated Reading List in Chicago Sun-Times Recommends Nonexistent Books, The New York Times, May 21, \url{https://www.nytimes.com/2025/05/21/business/media/chicago-sun-times-ai-reading-list.html}, retrieved August 21, 2025

Oransky, I., \& Marcus, A. (2010). Why write a blog about retractions? Retraction Watch, August 3. \url{https://retractionwatch.com/2010/08/03/why-write-a-blog-about-retractions}, retrieved August 20, 2025

Retraction Watch. (2024). Exclusive: Publisher retracts more than 450 papers from journal it acquired last year, August 23. \url{https://retractionwatch.com/2024/08/23/exclusive-publisher-retracts-more-than-450-papers-from-journal-it-acquired-last-year/}, retrieved April 5, 2025

Retraction Watch. (2025a). Paper retracted after author told journal study was ‘not actually performed’, January 24. \url{https://retractionwatch.com/2025/01/24/paper-retracted-after-author-told-journal-study-was-not-actually-performed/}, retrieved April 5, 2025

Retraction Watch. (2025b). The Retraction Watch Leaderboard, August 26. \url{https://retractionwatch.com/the-retraction-watch-leaderboard/}, retrieved August 27, 2025

Retraction Watch. (2025c). Dozens of Elsevier papers retracted over fake companies and suspicious authorship changes, May 14. \url{https://retractionwatch.com/2025/05/14/dozens-of-elsevier-papers-retracted-over-fake-companies-and-suspicious-authorship-changes/}, retrieved May 17, 2025

Retraction Watch. (2025d). Frontiers to retract 122 articles, links thousands in other publishers’ journals to “unethical” network, July 29. \url{https://retractionwatch.com/2025/07/29/frontiers-retract-122-articles-links-thousands-other-publishers-journals-to-unethical-network/}, retrieved August 3, 2025

Ryle, G. (1949). The Concept of Mind. University of Chicago Press

Sanger, L. (2004). Why Wikipedia must jettison its anti-elitism, December 31. \url{https://larrysanger.org/2004/12/why-wikipedia-must-jettison-its-anti-elitism/}, retrieved August 25, 2025

ScholarlyWorld. (2025). Springer Nature Retracts Nearly 3,000 Papers in 2024 Amid Research Integrity Efforts, February 24. \url{https://scholarlyworld.com/springer-nature-retracts-nearly-3000-papers-in-2024-amid-research-integrity-efforts/}, retrieved August 29, 2025

Sinclair, M. (2025). ChatGPT Traffic Overtakes Wikipedia, June 9. \url{https://www.azoma.ai/insights/wikipedia-chatgpt-citations-and-traffic-growth}, retrieved August 17, 2025

Sunstein, C. R. (2001). Republic.com. Princeton, NJ: Princeton University Press

Sweetland, J. H. (1989). Errors in bibliographic citations: A continuing problem. Library Quarterly, 59(4), 291–304

Temperton, J. (2015). Wikipedia’s world view is skewed by rich, western voices, September 15. \url{https://www.wired.com/story/wikipedia-world-view-bias/}, retrieved August 20, 2025

Tran, N. (2024). The ‘publish or perish’ mentality is fuelling research paper retractions – and undermining science. International Science Council, October 3. \url{https://council.science/blog/publish-or-perish-mentality/}, retrieved April 5, 2025

UnHerd. (2021). Wikipedia co-founder: I no longer trust the website I created, July 14. \url{https://www.youtube.com/watch?v=l0P4Cf0UCwU}, retrieved August 17, 2025

Vosoughi, S., Roy, D., \& Aral, S. (2018). The spread of true and false news online. Science, 359(6380), 1146–1151. \url{https://doi.org/10.1126/science.aap9559}

Wikipedia. (2023). Academic studies about Wikipedia. \url{https://en.wikipedia.org/wiki/Academic_studies_about_Wikipedia}, retrieved August 21, 2025

Wikipedia. (2005). Wikipedia: Press releases/Nature compares Wikipedia and Britannica, December 15. \url{https://en.wikipedia.org/wiki/Wikipedia:Press_releases/Nature_compares_Wikipedia_and_Britannica}, retrieved August 27, 2025

Wikipedia. (2024). Wikipedia: External peer review/Nature December 2005, April 1. \url{https://en.wikipedia.org/wiki/Wikipedia:External_peer_review/Nature_December_2005}, retrieved August 27, 2025

Wikipedia. (2025a). Citation needed, August 9. \url{https://en.wikipedia.org/wiki/Citation_needed}, retrieved August 21, 2025

Wikipedia. (2025b). Wikipedia: Don’t cite Wikipedia on Wikipedia, March 13. \url{https://en.wikipedia.org/wiki/Wikipedia:Don%27t_cite_Wikipedia_on_Wikipedia}, retrieved August 21, 2025

Wikipedia. (2025c). Wikipedia: Verifiability, August 22. \url{https://en.wikipedia.org/wiki/Wikipedia:Verifiability#Reliable_sources}, retrieved August 20, 2025

Wikipedia. (2025d). Wikipedia:Reliable sources, August 14, \url{https://en.wikipedia.org/wiki/Wikipedia:Reliable_sources}, retrieved August 23, 2025

Zhang, M., Wu, L., Yang, T., Zhu, B., \& Liu, Y. (2024). RETRACTED: The three-dimensional porous mesh structure of Cu-based metal-organic-framework – Aramid cellulose separator enhances the electrochemical performance of lithium metal anode batteries. Surfaces and Interfaces, 46, 104081. \url{https://doi.org/10.1016/j.surfin.2024.104081}

\end{document}